\documentclass[twoside,leqno,twocolumn]{article}

\usepackage[utf8]{inputenc}
\usepackage{ltexpprt}
\usepackage{hyperref}

\usepackage{graphicx}
\usepackage{amsmath}
\usepackage{mathtools} % for paired delim
\usepackage[binary-units]{siunitx} % for \SI
\usepackage[algo2e,ruled,vlined,linesnumbered]{algorithm2e}
\usepackage{booktabs}
\usepackage{subcaption}
\usepackage{csquotes}
\usepackage{thmtools}
\usepackage{thm-restate}

\DeclarePairedDelimiter\setSize{\lvert}{\rvert}

\NewDocumentCommand{\hypergraph}{}{\mathcal{F}}
\NewDocumentCommand{\vertexEdges}{m}{\hypergraph(#1)}
\NewDocumentCommand{\vertices}{}{V}
\NewDocumentCommand{\hypergraphSize}{}{\lVert \hypergraph \rVert}

\NewDocumentCommand{\hittingSet}{}{H}

\NewDocumentCommand{\edge}{}{F}
\NewDocumentCommand{\vertex}{}{v}

\graphicspath{{./graphics/}}
\begin{document}

\newcommand\relatedversion{}

\title{\Large An Efficient Branch-and-Bound Solver for Hitting Set\relatedversion}
\author{Thomas Bl\"asius \thanks{Karlsruhe Institute of Technology.}
\and Tobias Friedrich \thanks{Hasso Plattner Institute.}
\and David Stangl \footnotemark[2]
\and Christopher Weyand \footnotemark[1]
}

\date{}

\maketitle

% Copyright Statement
% When submitting your final paper to a SIAM proceedings, it is requested that you include
% the appropriate copyright in the footer of the paper.  The copyright added should be
% consistent with the copyright selected on the copyright form submitted with the paper.
% Please note that "20XX" should be changed to the year of the meeting.

% Default Copyright Statement
%\fancyfoot[R]{\scriptsize{Copyright \textcopyright\ 2022 by SIAM\\
%Unauthorized reproduction of this article is prohibited}}

% Depending on which copyright you agree to when you sign the copyright form, the copyright
% can be changed to one of the following after commenting out the default copyright statement
% above.

%\fancyfoot[R]{\scriptsize{Copyright \textcopyright\ 20XX\\
%Copyright for this paper is retained by authors}}

%\fancyfoot[R]{\scriptsize{Copyright \textcopyright\ 20XX\\
%Copyright retained by principal author's organization}}

%\pagenumbering{arabic}
%\setcounter{page}{1}%Leave this line commented out.

\begin{abstract} \small\baselineskip=9pt%
  The hitting set problem asks for a collection of sets over a
  universe $U$ to find a minimum subset of $U$ that intersects each of
  the given sets.  It is NP-hard and equivalent to the problem set
  cover.  We give a branch-and-bound algorithm to solve hitting set.
  Though it requires exponential time in the worst case, it can solve
  many practical instances from different domains in reasonable time.
  Our algorithm outperforms a modern ILP solver, the state-of-the-art
  for hitting set, by at least an order of magnitude on most
  instances.
  % the following empty line is necessary to get the spacing right...
  
\end{abstract}

%\clearpage

\section{Introduction}
\label{sec:intro}

Hitting set
naturally emerges from many problems appearing in various domains,
e.g., transportation~\cite{Weihe1998}, model-based diagnosis
\cite{Reiter1987}, data profiling~\cite{Birnick2020}, or
biology~\cite{itk-dritp-00}.  Unfortunately, hitting set is NP-hard.
In fact, it is among the first 21 NP-complete problems~\cite{Karp1972}.

Beyond its NP-completeness, there is a wide range of theoretic results
on hitting set, including exact algorithms~\cite{Shi2010},
approximation results~\cite{s-tagasc-97,cv-iaagsc-07,ds-aapr-14},
parameterized algorithms~\cite{a-kahs-10,f-pahs-06}, and parameterized
approximation algorithms~\cite{bf-paahs-12}.  Moreover, several
variants of the problem have been studied, e.g., weighted
variants~\cite{f-pahs-06}, geometric variants, where the instance
represents geometric objects~\cite{cv-iaagsc-07}, implicit hitting
set, where the instance is not explicitly given but implicitly by an
oracle that reveals sets not yet hit~\cite{ckmv-aihsp-11,mk-ihsas-13},
and the enumeration variant, where one has to find all inclusion-wise
minimal hitting sets instead of just the minimum~\cite{Gainer2017}.

Due to its importance for various applications, hitting set has also
been studied from a practical perspective.  A lot of engineering work
has been dedicated to the above mentioned enumeration variant; see the
survey by Gainer-Dewar and Vera-Licona~\cite{Gainer2017} for an
overview and the paper of Murakami and Uno~\cite{Murakami2014} for the
state-of-the-art algorithm.  For the optimization problem of finding a
minimum hitting set, there are results on heuristic algorithms, e.g.,
\cite{b-gamhs-14,ckw-scavld-10}, as well as heavily parallelized
brute-force approaches using GPUs~\cite{ccm-fehss-17,cym-mghsagi-15}.

Concerning clever algorithmic techniques for solving hitting set
exactly, there is the seminal work of Weihe~\cite{Weihe1998} proposing
two rules for data reduction that perform very well on instances
coming from rail networks~\cite{blaesius2019}.  More recently, Bevern and
Smirnov~\cite{Bevern2020} proposed alternative reduction rules for
$d$-hitting set (restricting the size of each set to at most $d$) and
evaluated them on instances coming from the cluster vertex deletion
problem.  Though reduction rules are a crucial component in designing
efficient algorithms, one generally still needs an algorithm to solve
the remaining instance.  Concerning such an algorithm, the current
state-of-the-art is somewhat unsatisfactory.  In 2000, Caprara, Toth,
and Fischetti~\cite{Caprara2000} did an exhaustive study of all
prevalent solvers at the time and conclude:
\blockquote[\cite{Caprara2000}]{This shows that the state-of-the-art
  general-purpose ILP solvers are competitive with the best exact
  algorithms for SCP\footnote{SCP stands for ``set cover problem'',
    which is equivalent to the hitting set problem.}  presented in the
  literature, and that their performance can sensibly be improved by
  an external preprocessing procedure.}
Later, de Kleer~\cite{Kleer2011} conducted an empirical study on the
effect of Weihe's reduction rules~\cite{Weihe1998} in a simple
branch-and-bound algorithm.  However, the algorithm does not
outperform a general-purpose ILP solver.  To the best of our
knowledge, using an ILP solver, potentially after preprocessing,
remains the state-of-the-art to this day.

In this paper, we engineer and evaluate a branch-and-bound algorithm
that beats this state-of-the-art.
% see Figure~\ref{fig:vs-gurobi}.
On our test set of $929$ instances where the ILP solver reported a
non-zero\footnote{Gurobi
  reports running times below $0.01$ as $0$.}  running time, we reach
a median speedup factor of more than $25$.  For three quarters of
these instances, we have a speedup of more than one order of
magnitude.

The basic building blocks of our branch-and-bound algorithm are bounds
on the solution size and data reduction rules.  They are described in
Section~\ref{sec:hs}.  We note that most bounds and reduction rules we
use have been considered before, either for hitting set or in a
different context.  For the different lower bounds we give a
theoretical analysis that completely characterizes how they relate to
each other; see Section~\ref{sec:lower-bounds}.  In
Section~\ref{sec:algo} we specify our overall algorithm and provide
details on how to efficiently implement it.  Our evaluation in
Section~\ref{sec:eval} is based on $4256$ instances from
different domains.  Beyond the overall running time of our algorithm,
we give a detailed evaluation of how much the different building
blocks contribute to the final result.  Our implementation is publicly 
%available.
available\footnote{\url{https://github.com/Felerius/findminhs}}.

%%% Local Variables:
%%% mode: latex
%%% TeX-master: "../paper"
%%% End:

\section{Basic Building Blocks}
\label{sec:hs}

In this section we describe lower and upper bounds as well as reduction rules and introduce the hitting set problem and our notation.

\subsection{Problem Definition.}

Formally, a \emph{hypergraph} $\hypergraph$ is a set family over a \emph{vertex set} $\vertices$, i.e., every $\edge \in \hypergraph$ is a subset of $\vertices$.
We call these elements $\edge \in \hypergraph$ \emph{hyperedges}.
For brevity, we will refer to them as \emph{edges}.
The number of vertices in the hypergraph is $\setSize{\vertices}$, and likewise the number of edges $\setSize{\hypergraph}$.
Additionally, we use $\hypergraphSize$, called the hypergraph size, to refer to the sum of all edge sizes, i.e., $\hypergraphSize = \sum_{\edge \in \hypergraph} \setSize{F}$.
For a vertex $\vertex \in \vertices$, we denote the set of edges containing $\vertex$ as $\vertexEdges{\vertex}$.
We call $\deg(\vertex) = \setSize{\vertexEdges{\vertex}}$ the \emph{degree} of $\vertex$.
Note that the sum of vertex degrees is equal to $\hypergraphSize$.

% We also define the \emph{associated bipartite graph} of a hypergraph $\hypergraph$.
% It contains one vertex for every vertex and edge in the hypergraph.
% Two vertices in it are connected by an edge if and only if one represents a vertex $\vertex \in \vertices$, the other an edge $\edge \in \hypergraph$, and $\vertex \in \edge$.

We say that a vertex \emph{hits} an edge if it is contained in it.
Based on this, we call a vertex subset $\hittingSet \subseteq \vertices$ a \emph{hitting set} of $\hypergraph$ if all edges in $\hypergraph$ are hit by at least one vertex in $\hittingSet$.
Formally, $\hittingSet \subseteq \vertices$ is a hitting set of $\hypergraph$ if and only if $\forall \edge \in \hypergraph: \hittingSet \cap \edge \neq \emptyset$.
We call a hitting set \emph{minimum} if no smaller hitting set for the same hypergraph exists.
We refer to a hitting set as \emph{minimal} if it contains no other hitting set as proper subset.
The hitting set problem asks for a minimum hitting set of a given hypergraph.
%The set of minimal hitting sets of a hypergraph form its dual, which is itself a hypergraph.
%The dual of the dual is the original hypergraph.

\subsection{Upper Bounds.}
\label{sec:upper-bounds}

For the upper bound, we use the simple greedy algorithm of repeatedly picking the vertex with the highest degree.  This results in a $\log n$-approximation, works well in practice, and runs in linear time~\cite{Grossman1997,Skiena2020}.
We note that there are multiple LP-based upper (and also lower) bounds for which we refer to the overview by Caprara~\cite{Caprara2000}.  

\subsection{Lower Bounds.}
\label{sec:lower-bounds}

In contrast to upper bounds, good lower bounds are harder to achieve, but crucial for the pruning.  Here we describe five lower bounds, some of which have been used for hitting set or other problems.  
Moreover, we prove a complete characterization of how the lower bounds
relate to each other.

\begin{description}
\item[max-degree bound]
The max-degree bound uses that each vertex hits at most $d_{\max}$ many edges, where $d_{\max}$ is the highest vertex degree. 
Thus, at least $\left\lceil\frac{\setSize{\hypergraph}}{d_{\max}} \right\rceil$ vertices are required to hit all edges.
\item[sum-degree bound] 
Let $d_1, \dots, d_n$ be the vertex degrees in descending order.
Since vertices can only be chosen once, the max-degree bound can be improved to the smallest $k$ for which 
$\sum_{i=1}^k d_i \geq \setSize{\hypergraph}$.
\item[efficiency bound] 
Consider any solution $S$. 
Let each vertex $\vertex\in S$ charge its cost onto the edges it hits.
That is, each edge $F\in \vertexEdges{\vertex}$ is charged $1/\deg(\vertex)$ by $\vertex$.
The size of the solution can now be expressed as the sum of the cost of all edges, 
i.e.,~ $|S|=\sum_{\edge\in\hypergraph} \sum_{\vertex\in S\cap\edge} 1/\deg(\vertex)$.
%The degree $\deg(\vertex)$ of each vertex can be seen as the efficiency of a vertex when hitting edges.
The efficiency bound assumes the lowest cost for each edge individually, yielding
$\left\lceil \sum_{\edge\in\hypergraph} \min_{\vertex\in\edge} \frac{1}{\deg(\vertex)} \right\rceil$ as a lower bound.
\item[packing bound]
A set $P$ of pairwise disjoint edges constitutes a lower bound, because each vertex appears in at most one of those edges.
Thus at least $|P|$ vertices are required to cover them.
Finding the best packing bound is actually an independent-set problem on the intersection graph of the edges $\hypergraph$.
Using an independent set of conflicts as a lower bound is a known technique applied by recent solvers for other hard problems~\cite{Gottesbueren2020}.
\item[sum-over-packing]
Any packing $P$ can be used to strengthen the sum-degree bound, as the packing requires to select $|P|$ vertices to cover the edges in $P$, which might not be the vertices of highest degree that the sum bound would use otherwise.  In the remainder, we focus on how many edges of $\hypergraph \setminus P$ we cover.
The vertices selected to cover $P$ can cover at most $b_P = \sum_{\edge\in P} \max_{\vertex\in\edge} (\deg(\vertex) - 1)$ edges in $\hypergraph \setminus P$.
If $b_P$ is smaller than $\setSize{\hypergraph \setminus P}$, we have to pick additional vertices to cover the remaining $\setSize{\hypergraph \setminus P} - b_p$  edges.
To this end, let $d_1, \dots, d_n$ be the descending vertex degrees in $\hypergraph \setminus P$, excluding the vertex of highest degree of each edge in $P$ (these vertices have already been selected and cannot be selected twice).
With this, the improved sum bound is $|P| + k$ where $k \geq 0$ is the smallest number for which $\sum_{i=1}^k d_i \geq \setSize{\hypergraph \setminus P} - b_P$.
\end{description}

\begin{figure}
    \centering
    \includegraphics{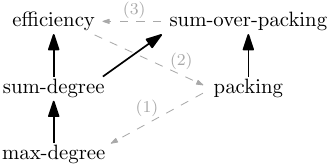}
    \caption{
    Relations between lower bounds. 
    A bold arrow from $a$ to $b$ means that $a$ is dominated by $b$ (Lemma~\ref{lem:domination}). 
    Dashed arrows are labeled as in Lemma~\ref{lem:incomp} and indicate that there exists an instance where $a$ is smaller than $b$.  Note that for any pair of bounds, there is either a directed bold path (indicating dominance) or a directed path containing exactly one dashed edge (indicating non-dominance).
    }
    \label{fig:lower-hierarchy}
\end{figure}

A lower bound $a$ dominates another bound $b$, if $a\geq b$ for all problem instances.
If a bound has multiple choices (e.g.,~the packing bounds), we
consider the choice that leads to the highest bound.
%For sum-over-packing, that does not necessarily have to be a maximum packing.
Two bounds are incomparable if neither dominates the other, i.e.,~there is an instance where $a<b$ and another instance where $a>b$.
Figure~\ref{fig:lower-hierarchy} shows the relations between the lower bounds stated in the following lemmas.
%Their proofs can be found in section~\ref{sec:proofs}.

% domination
\begin{lemma}
\label{lem:domination}
The sum-over-packing dominates the sum-degree and the packing bound.
The sum-degree bound dominates the max-degree bound and is dominated by the efficiency bound.
\end{lemma}
\begin{proof}
We first show that the sum-over-packing bound dominates the sum-degree, and the packing bound.
The packing bound cannot be larger because the sum-over-packing bound adds a non-negative number to it.
%The sum-degree bound cannot be larger, because the sum-over-packing bound can be seen as a sum-degree bound where the sum is forced to include smaller degrees, thus requiring more elements.
Also the sum-over-packing bound over an empty packing is exactly the sum-degree bound.

The max-degree bound can be expressed as the smallest $k$ for which $\sum_{i=1}^k d_1 \geq \setSize{\hypergraph}$ 
showing that the sum-degree bound dominates the max-degree bound.

It remains to show that the efficiency bound dominates the sum-degree bound.
The sum-degree bound is given by the $k$ highest degree vertices whose degrees sum up to just above $\setSize{\hypergraph}$.
Let $\vertex_1,\dots,\vertex_k$ be these vertices sorted descending by degree.
%
% attempt to make it easier
We partition the edges of $\hypergraph$ into $k$ sets
$E_1, \dots, E_k$ with the following two properties.  First,
$|E_i| = \deg(\vertex_i)$ for $i < k$ and
$1 \le |E_k| \le \deg(\vertex_k)$.  Second, the maximum vertex degree
for every edge in $E_i$ is at most $\deg(\vertex_i)$.  Such a
partition can be achieved as follows.  Assign to $E_1$ all edges
containing $\vertex_1$.  For larger $i$, assign to $E_i$ all edges
that contain $\vertex_i$ that have not yet been assigned.  Moreover,
add further edges arbitrarily, until $E_i$ contains $\deg(\vertex_i)$
edges.  The first property, concerning the sizes of $E_i$, clearly
holds.  For the second property, observe that in step $i$, all edges
containing vertices $v_1, \dots, v_{i - 1}$ have already been
assigned.  Thus, all unassigned edges, and thereby all edges ending up
in $E_i$, only contain vertices of degree at most $\deg(\vertex_i)$.

With this partition, we get that the efficiency bound is larger than $k-1$, because 
\begin{align*}
\sum_{F\in \hypergraph} \min_{\vertex\in\edge} \frac{1}{\deg(\vertex)} 
&= \sum_{i=1}^{k}\sum_{F\in E_i} \min_{\vertex\in\edge} \frac{1}{\deg(\vertex)} \\
&\geq \sum_{i=1}^{k}\sum_{F\in E_i} \frac{1}{\deg(\vertex_i)}\\
%= \sum_{i=1}^{k} \frac{|E_i|}{\deg(\vertex_i)}
&= k-1 + \frac{|E_k|}{\deg(\vertex_k)} 
> k-1,
\end{align*}
and thus the rounded-up efficiency bound is at least as high as the sum-degree bound.
%$$\left\lceil \sum_{F\in \hypergraph} \min_{\vertex\in\edge} \frac{1}{\deg(\vertex)} \right\rceil \geq k.$$
\end{proof}

\begin{figure}
    \centering
    \includegraphics{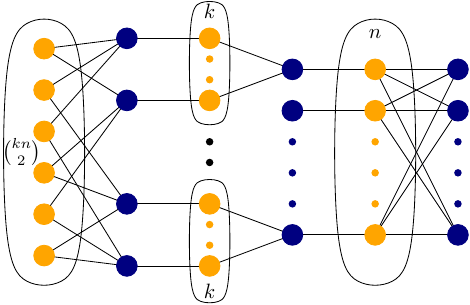}
    \caption{Instance family represented as a bipartite graph. The sum-over-packing bound is worse than efficiency for $k=2$ and efficiency $<$ packing for $k=1$. Edges are blue and vertices are orange.}
    \label{fig:lower-example}
\end{figure}

%incomparability
\begin{lemma}
\label{lem:incomp}
The packing bound is incomparable with the max-degree, sum-degree, and efficiency bound.
The efficiency bound is incomparable with the sum-over-packing bound.
\end{lemma}
\begin{proof}
We give examples of instances where (1) packing $<$ max-degree, (2) efficiency $<$ packing, and (3) sum-over-packing $<$ efficiency; see Figure~\ref{fig:lower-hierarchy}.
%(1) packing<max (triangle)\\
%(2) efficiency < packing (construction with k=1)\\
%(3) sum-over-packing < efficiency (construction with k=2)\\
With the domination relations from Lemma~\ref{lem:domination}, the incomparability of the packing bound follows from the instances where 
packing $<$(1) max-degree $\leq$ sum-degree $\leq$ efficiency $<$(2) packing.
Similarly, the second part of the lemma follows from the instances where efficiency $<$(2) packing $\leq$ sum-over-packing $<$(3) efficiency.

% show 1
In the example for (1), there are three vertices $a,b,c$ and three edges $\{a,b\}, \{a,c\}, \{b,c\}$.
Every packing contains at most one edge.
In contrast the max-degree bound is $\lceil 3/2 \rceil = 2$.

% construction for 2,3
We give a parametrized example for (2) and (3).
The construction is shown in Figure~\ref{fig:lower-example}.
There are three types of edges.
In the center, there are $n$ disjoined edges, the \emph{center edges}, that constitute a maximum packing.
To the right there are $n$ edges, the \emph{right edges}, that share $n$ vertices.
To the left are $k\cdot n$ edges, the \emph{left edges}, that pairwise share a vertex of degree two.
Each of the center edges contains one of $n$ high degree vertices shared by the right edges.
Each center edge also contains $k$ vertices of degree two that are also contained in one of the $k\cdot n$ left edges.

% show 2
For $k=1$, the efficiency bound is (for edges from left to right) $n/2 + n/(n+1) + n/(n+1)<n$ while the largest packing, e.g., the center edges, has size $n$.
Thus, the efficiency bound can be smaller than the largest packing showing (2).

% show 3
For $k=2$ the efficiency bound is $2n/2 + n/(n+1) + n/(n+1) > n$.
To prove (3), it remains to show that the sum-over-packing bound is at most $n$ for each possible packing.
Note that the argument must hold for \emph{each} packing and not just maximum packings because a maximum packing does not necessarily lead to the highest sum-over-packing bound.
If a packing includes a right edge, this prevents the inclusion of all other right edges and all center edges.
If a packing includes a left edge, this prevents the inclusion of all other left edges and exactly one center edge.
The only disjoined packings possible are either only center edges, one left edge and the rest center edges, or one left and one right edge.
Thus, each packing $P$ has size at most $n$ and contains at least $|P|-2$ center edges, which have vertices of degree $n+1$.

Now we show that the sum-over-packing bound is either constant or limited by the packing size.
For sufficiently large $n$ and any packing $S$ that is larger than six, the packing contains at least four center edges.
The sum-over-packing bound adds nothing to the size of the packing because the maximum degrees of the vertices in four center edges add up to more than $\setSize{\hypergraph}$.
That is, the sum-over-packing bound is $|S|$, because $\sum_{\edge\in S} \max_{\vertex\in\edge} \deg(\vertex) > 4n = \setSize{\hypergraph}$.
For any packing $S$ smaller than six, the remaining instance after deleting $S$ and its max-degree vertices
still contains $\Omega(n)$ nodes of degree at least $n$. 
Four of them suffice to obtain a degree sum larger than $\setSize{\hypergraph}$.
Therefore the sum-over-packing bound is at most $|S|+4 \leq 10$.
In each case, the sum-over-packing bound is at most $n$, thus smaller than the efficiency bound.
\end{proof}

\subsection{Reduction Rules.}

Our algorithm uses the following reduction rules.  The domination rules were first described by Weihe~\cite{Weihe1998}.  The costly discard rule is a standard technique in branch and bound but has, to the best of our knowledge, not yet been used for hitting set.  The unit edge rule is widely known and stated, e.g., by Shi and Cai~\cite{Shi2010}.

\begin{description}
\item[Unit Edge Rule.] If there is an edge of size one, then pick the contained vertex.
\item[Edge Domination Rule.] If there are two edges $e_1, e_2$ with $e_1 \subseteq e_2$, then delete $e_2$.
\item[Vertex Domination Rule.] If there are two vertices $v_1,v_2$ such that $\vertexEdges{v_1} \subseteq \vertexEdges{v_2}$, then delete $v_1$.
\item[Costly Discard Rule.] If discarding a vertex raises the lower bound to or above the current upper bound, then pick this vertex.
\end{description}

We do not consider the complement of the costly discard rule (costly inclusion) because including a vertex cannot raise the packing lower bound by more than one.
The rule could only take effect when the upper and lower bound differ by one, in which case the instance is almost solved anyway.

We note that Shi and Cai~\cite{Shi2010} proved that branch-and-bound runs in time $O(1.23801^n)$ when using the first three of the above reduction rules in addition to two rules based on edges of size two.  We omitted these two rules from our solver, as they rarely take effect.

%%% Local Variables:
%%% mode: latex
%%% TeX-master: "../paper"
%%% End:

\section{The Branch-and-Bound Algorithm}
\label{sec:algo}

In this section we describe the structure and efficient implementation of our solver as outlined in Algorithm~\ref{alg:solver}.
In every step, our algorithm branches on the inclusion or exclusion of a vertex in the solution, thereby creating two new instances that are solved recursively.
Before each branch, we apply the follow two steps.
% The recursive subroutine proceeds in three steps.
First, an approximate solution for the current instance is found using a greedy algorithm.
The best found solution so far represents the current upper bound and is globally maintained as a result of the algorithm.  It is used to prune branches where no better solution can be achieved. 
Second, reduction rules are repeatedly applied until either no reduction is possible or a lower bound allows the current branch to be pruned.

\begin{algorithm2e}
\DontPrintSemicolon
update upper bound \tcp*[l]{greedy}
\While{reductions or pruning possible}{
    compute lower bounds\;
    \lIf{lower bound $\geq$ upper bound}{\KwRet{}}
    apply first applicable reduction
}
select branching vertex $v$ by highest degree\;
branch on choosing $v$ \tcp*[l]{inclusion branch}
branch on discarding $v$ \tcp*[l]{exclusion branch}
\caption{Solve Recursively.}
\label{alg:solver}
\end{algorithm2e}

\subsection{Operation Summary and Reduction Order.}
\label{sec:algo-order}
Initially we compute a greedy upper bound in time $O(||\hypergraph||)$ but do not repeat it in the reduction loop.
Then, reductions and bounds are checked one after another (Algorithm~\ref{alg:solver}, line 2-5).
They are processed in ascending order by runtime to prevent expensive operations when possible.
After one reduction is applied, the search starts from the top of the list again.
In general, lower bound pruning happens before reductions.
Although the max-degree bound is dominated by the efficiency bound, we include it due to the lower computational complexity.
%Contrasting this, we do not use the sum-degree bound because it has just the same running time as the efficiency bound.
The order of lower bounds and reductions is as follows.

\begin{enumerate}
\item max-degree bound in $O(|V|)$
\item efficiency bound in $O(||\hypergraph||)$
\item packing bound in $O(||\hypergraph|| + |\hypergraph|\log|\hypergraph|)$
\item sum-over-packing bound in $O(||\hypergraph||)$
\item unit edge rule in $O(|\hypergraph|)$
\item costly discard rule with efficiency bound updates for all vertices in $O(||\hypergraph||)$
\item costly discard rule with packing updates for all vertices in $O(||\hypergraph|| + |\hypergraph|\log|\hypergraph|)$
\item costly discard rule with repack for $3$ vertices in $O(||\hypergraph|| + |\hypergraph|\log|\hypergraph|)$
\item edge domination rule in $O(|\hypergraph|\cdot||\hypergraph||)$
\item vertex domination rule in $O(|V|\cdot||\hypergraph||)$
\end{enumerate}

In the following, we discuss the branching strategy and implementation
details of the instance representation, the bound computation, and the reduction rules.

\subsection{Branching Strategy.}
As mentioned above, we branch on the inclusion or exclusion of a verticex.
The remaining degrees of freedom are the vertex to branch on and the order in which the two branches are processed.
We found the latter to be irrelevant while the former crucially affects search space and performance.
Our solver always branches on the vertex with highest degree in the remaining instance and processes the inclusion branch first.

% We note that it is also possible to branch on edges.  This has been done e.g.~by algorithms to enumerate all minimal hitting sets~\cite{Murakami2014}.  
% To find the minimum hitting set, however, a simpler and more effective strategy is to branch on the inclusion or exclusion of a node in the solution~\cite{Shi2010,Kleer2011}.

% activity
% A technique that is applied to great success in solvers for the Boolean satisfiability problem (SAT)~\cite{Balyo2020} 
% is the \emph{Variable State Independent Decaying Sum} (VSIDS) heuristic~\cite{Moskewicz2001}. 
% Requiring little problem-specific knowledge, it determines the branching node based on so called \emph{activity values} that represent importance of nodes.
% Activity values are updated while solving the instance, meaning the solver gradually learns the structure of the given instance with the goal of pruning the search tree early and often.
% Preliminary experiments against a uniform random choice showed promising results when using VSIDS in our hitting-set solver.\todo{cite theses?}
% The tests revealed that the VSIDS heuristic produces activity values that strongly correlate with node degree.
% In fact, always choosing the node with the highest degree achieves slightly better results and is more consistent. 
% Nevertheless, it's interesting that VSIDS approximates the max-degree strategy.
% It shows that the concept is capable of \emph{learning} a good strategy and might be applicable to other problems.

\subsection{Instance Representation.}
\label{sec:inst-rep}
% changes and rollbacks for recursion
During the algorithm the instance needs to be updated regularly.  To
avoid copying the instance, we maintain one data structure
representing the current instance throughout the algorithm.  There are
three places in the algorithm where the instance is modified.  First,
the greedy algorithm iteratively deletes vertices.  Second, the
reduction rules reduce the instance.  Third, when branching, a vertex is
excluded (it is deleted) or is included (it and all its edges are
deleted).  In all cases, changes have to be rolled back appropriately.

% sets must be sorted
To support these operations, we maintain the vertices of each edge and the edges of each vertex in sorted order at all times.
Maintaining the order speeds up set-like operations, e.g.,~union of two edges, and is required by the reduction rules for edge and vertex domination.

For this, we implement a data structure called \emph{ordered subset
  list} that manages a subset $S\subseteq \{s_1,\dots,s_n\}$ of $n$
strictly-ordered objects $s_1 < s_2 < \cdots < s_n$. 
Assuming the $s_i$ are sorted in advance, it supports the following operations. 
\begin{center}
\begin{tabular}{r l c}
  \toprule
    Name & Description & Time \\
  \midrule
    $\operatorname{init}()$ & Initialize $S=\{s_1,\dots,s_n\}$ & $O(n)$ \\
    $\operatorname{del}(i)$ & Delete $s_i$ from $S$ & $O(1)$ \\
    $\operatorname{undo}()$ & Undo last (not undone) deletion & $O(1)$ \\
    $\operatorname{iter}()$ & Traverse $S$ in increasing order & $O(\setSize{S})$ \\
    $\operatorname{iterrev}()$ & Traverse $S$ in decreasing order & $O(\setSize{S})$ \\
  \bottomrule
\end{tabular}
\end{center}
This can be implemented by storing the subset $S$ itself in a
doubly-linked list.  Additionally, we have an array $A$ that points
for each $i$ to the list entry corresponding to $s_i$.  When deleting
an element from $S$, its list entry can be found in constant time via
$A$.  It is removed from the list, but the list item itself remains in
memory and $A$ keeps the pointer to it.  To allow for later
reinsertion, we maintain a stack of indices of deleted list entries.
As the list entry itself has not been modified at the time of
deletion, its previous and next entries are intact and thus we can
reinsert it into the list in constant time in the position it was
before its deletion.

\subsection{Upper Bound Computation.}
The greedy algorithm picks the highest degree vertex and deletes it and its edges from the instance until all edges are hit.
Since modifications, i.e.,~deleting edges and vertices, are done in time linear in the number of changes (see Section~\ref{sec:inst-rep}), this totals to at most linear time until a hitting set is found.
Finding the vertex with highest degree in each step is done with a bucket heap~\cite{Cormen2009, Skiena2020} that stores the vertex degrees.
The data structure allows constant time operations due to the limited range of the stored values. 
Since degrees are only lowered during the procedure and the total vertex degree is $||\hypergraph||$ the greedy algorithm takes linear time in the size of the instance.

% A \emph{bucket heap} manages $n$ values $a_1$ to $a_n$ with $0 \leq a_i \leq m$.
% It requires $O(n + m)$ memory and supports the following operations:
% \begin{center}
% \begin{tabular}{r l c}
%   \toprule
%     Name & Description & Time \\
%   \midrule
%     $\operatorname{decrement}(i)$ & Decrease $a_i$ by one & $O(1)$ \\
%     $\operatorname{max}()$ & Return $\max \{a_1, \dots, a_n\}$ & $O(1)$ \\
%   \bottomrule
% \end{tabular}
% \end{center}
% \begin{proof}
%   To support this data structure, we need a linked list $L_j$ for every possible value of $a_i$, i.e., from $L_0$ to $L_m$.
%   The list $L_j$ contains the indices $i$ for which $a_i = j$.
%   Additionally, an array of length $n$ references for all $i$ the list entry in $L_{a_i}$ corresponding to $a_i$.
%   A variable is kept equal to the maximum $a_i$, which is returned from $\operatorname{max}()$.
%   With this setup, $\operatorname{decrement}(i)$ moves a list entry from one list to another and possibly updates the maximum which changes by at most one.
%   %Since the maximum $a_i$ is changed by at most one, it can be updated in constant time.
% \end{proof}

\subsection{Packing Bound Computation.}
\label{sec:packing-computation}

%\paragraph*{Initial Packing}
Finding a maximum packing of disjoined edges is an independent set problem and thus computationally expensive.
Recent solvers for the quasi-threshold editing and cluster editing problem, which use the same idea of packing conflicts, apply the min-degree heuristic to find a good packing~\cite{Gottesbueren2020, Hartung2015}.
In our context, the degree in the conflict graph of an edge $\edge$ from the original instance would be the number of other edges that share at least one vertex with $\edge$.
We approximate this and sort all edges by $\sum_{\vertex\in\edge} \deg(\vertex)$ in ascending order.
Then, we go through the edges and add the current edge to the packing if possible.
When adding an edge to the packing, each contained vertex is marked.
An edge $\edge$ can be added if all contained vertices are unmarked, which can be checked in $|\edge|$.
In total, the initial packing is computed in $O(||\hypergraph|| + |\hypergraph|\log|\hypergraph|)$.

%\paragraph*{Packing Growth}
We implemented the 2-improvement heuristic for independent set to grow the packing~\cite{Andrade2012}.
The heuristic is a local search that repeatedly tries to replace an element from the packing with two new ones.
Although this technique is effective~\cite{Gottesbueren2020}, we found it to be too slow and too rarely applicable to justify the high computational cost (see Figure~\ref{fig:search-space-by-bound}).
Our implementation runs in $O(|P|\cdot||\hypergraph||)$ per improvement where $P$ is the current packing.

%\paragraph*{Efficiency bound}
%The efficiency bound is numerically challenging.
%We use 64-bit floating point numbers to add up the fractions.
%Before rounding, we first subtract a small epsilon to prevent
%a value that should be integral to be rounded up due to floating point inaccuracy.

\subsection{Efficient Costly Discard Rule.}

The costly discard rule states that a vertex must be picked if discarding it raises some lower bound to or above the current upper bound.
That is, if we were to branch on that vertex, the exclusion branch would be pruned immediately.
The rule has two degrees of freedom: first, the vertex it is applied to and, second, the lower bound that is used.
For maximum effectiveness of the reduction, we would like to check the rule for all vertices and lower bounds.
However, computing all lower bounds from scratch $|V|$ times is computationally expensive.
We restrict it to the efficiency and packing bound. In the following, we discuss how to compute these bounds efficiently for all vertices at once.

\paragraph{Costly Discard with Efficiency Bound.}
For the efficiency bound, checking the costly discard rule for all vertices at once can be done in $O(||\hypergraph||)$ as follows.
First, the efficiency bound is computed for the current instance.
While doing so, for each edge the two vertices with highest degree are saved.
When a vertex $\vertex$ would be discarded from the instance, only edges $\vertex$ is contained in can change their contribution to the bound.
Such an edge $\edge$ changes the contribution $\min_{u\in\edge}1/\deg(u)$ only if $\vertex$ was the vertex with highest degree in $\edge$.
In this case the contribution depends on the vertex with second highest degree in $\edge$, which was identified earlier.
In total, discarding $\vertex$ changes the contribution of at most $\deg(\vertex)$ edges that can be updated in constant time each.
Over all vertices this sums up to $||\hypergraph||$.

\paragraph{Costly Discard with Packing Bound.}
For the packing bound, a similar approach of dynamically updating a packing bound $|V|$ times can be used to check the rule for all vertices and constitutes item 7~in Section~\ref{sec:algo-order}.
Discarding a vertex $\vertex$ removes it from all edges.
The edges that are relevant are those that intersect the union of the current packing exactly in vertex $\vertex$.
That is, they could now be included in the packing after $\vertex$'s removal.
We say that such an edge is blocked by vertex $\vertex$.
Each edge is blocked by at most one vertex.
After the initial packing is constructed, we create for each vertex $\vertex$ a list of edges that are blocked by this vertex and sort each list individually by the highest degree of a contained vertex (excluding $\vertex$).
These lists are found and sorted in $O(||\hypergraph||+ |\hypergraph|\log|\hypergraph|)$.
When checking if a vertex qualifies for the costly discard rule with the packing bound, we traverse the list of blocked edges for this vertex and greedily add them to the packing if possible.
After the rule is checked, we remove the added edges to restore the initial packing.
Since each edge $\edge$ is in at most one list and can be added to the packing in time $O(|\edge|)$,
the costly discard rule can be checked for all vertices in time $O(||\hypergraph||)$ when using the lists.
The creation of the lists dominates the running time with $O(||\hypergraph|| + |\hypergraph|\log|\hypergraph|)$, which is the same as the time it takes to compute the initial packing.

Unfortunately, the updated packings are worse than if they were computed from scratch.
Therefore, we additionally choose the $c$ vertices of highest degree, for which we check the costly discard rule with a completely new packing each (see item 8~in Section~\ref{sec:algo-order}).  Our experiments in Section~\ref{sec:solver-details} suggest that $c = 3$ is a reasonable choice.

\subsection{Efficient Domination Rules.}

The domination rules can be checked naively by comparing each set with all others to find inclusions.
This implies a running time of $O(|V|\cdot||\hypergraph||)$  and $O(|\hypergraph|\cdot||\hypergraph||)$ which is quadratic in the number of edges or vertices, respectively.
In fact, under the strong exponential time hypothesis the reduction cannot be done in less than quadratic time in the worst case~\cite{borassi2016}.
In practice, however, the sub- and superset reductions can be sped up significantly using \emph{set tries}, a data structure described by Savnik~\cite{Savnik2013}.

A set trie manages a collection $\mathcal{T}$ of sets over $[n]$ and supports the following operations which require that the given sets are sorted and can be traversed in linear time.
\begin{center}
\begin{tabular}{r m{8em} c}
  \toprule
    Name & Description & Time \\
  \midrule
    $\operatorname{add}(S)$ & add $S$ to $\mathcal{T}$ & $O(\setSize{S})$ \\
    $\operatorname{hasSubset}()$ & does $\mathcal{T}$ contain a subset of $S$ & $O(\setSize{S} + \lVert \mathcal{T} \rVert)$ \\
    $\operatorname{hasSuperset}()$ & does $\mathcal{T}$ contain a superset of $S$ & $O(\setSize{S} + \lVert \mathcal{T} \rVert)$ \\
  \bottomrule
\end{tabular}
\end{center}

%\begin{description}
%\item[add(S)] add $S$ to $\mathcal{T}$ in 
%$\mathcal{O}(\setSize{S})$
%\item[hasSubset(S)] does $\mathcal{T}$ contain a subset of $S$ in 
%$\mathcal{O}(\setSize{S} + \lVert \mathcal{T} \rVert)$
%\item[hasSuperset(S)] does $\mathcal{T}$ contain a superset of $S$ in 
%$\mathcal{O}(\setSize{S} + \lVert \mathcal{T} \rVert)$
%\end{description}

% dominating edge (remove superset)
For the dominated edge rule we create an empty set trie and iterate through the edges in increasing order of their size.
For each edge, we check whether a subset of it exists in the set trie. 
If so, the edge is dominated. Otherwise the edge is added to the set trie. 
Note that this process can be continued after the first dominated edge to find all of them. 
% dominating vertex

For the dominated vertex rule, the process is similar. 
An empty set trie is created that stores sets of edges.
Then, vertices are iterated in decreasing order of their degree.
Recall that $F(v)$ is the set of edges containing the vertex $v$.
If the set trie contains a superset of $F(v)$, the vertex $v$ is dominated. 
Otherwise, $F(v)$ is added to the set trie. 
%The order in which sets are inserted into the trie is ascending by size for the superset reduction and descending for the subset reduction.

%%% Local Variables:
%%% mode: latex
%%% TeX-master: "../paper"
%%% End:

\section{Evaluation}
\label{sec:eval}

\begin{figure}
  \includegraphics[width=\linewidth]{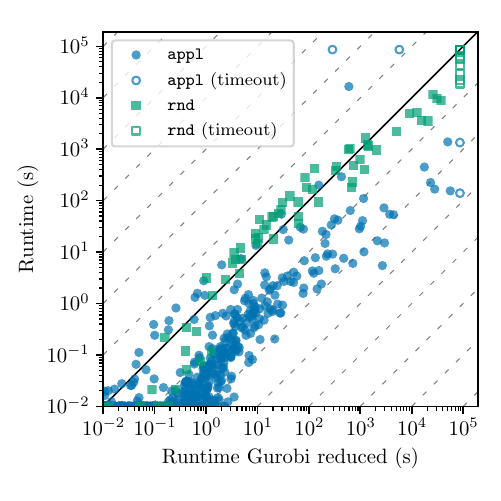}
  \caption{Run time of our solver compared to Gurobi with preprocessing. Times are rounded up to 0.01s.}
  \label{fig:vs-gurobi}
\end{figure}
\begin{figure*}
  \includegraphics[width=\textwidth]{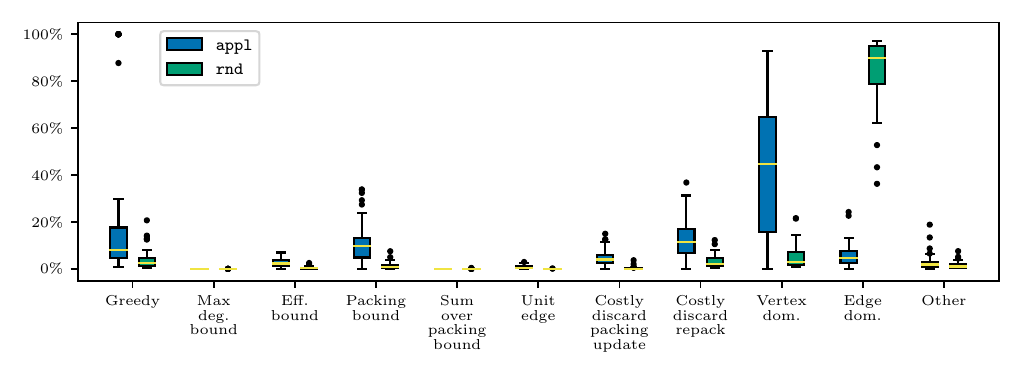}
  \caption{For each instance, the run time share of each operation was measured relative to the total run time of the instance. Note that the time for efficiency costly discard is included in the efficiency bound time.}
  \label{fig:operations}
\end{figure*}

In this section, we evaluate our branch-and-bound algorithm experimentally.  First, in Section~\ref{sec:runtime}, we compare its run time to the state-of-the-art ILP-solver Gurobi~\cite{gurobi} and analyze how much time is spent in which part of the algorithm.
%Then, in Section~\ref{sec:search-space-bounds}, we examine the behavior of our algorithm in detail by studying the size of the search space and the upper and lower bounds we obtain.
In Section~\ref{sec:solver-details}, we investigate details regarding the performance and effectiveness of lower bounds, upper bounds, and reduction rules to evaluate the used techniques as well as to substantiate our design decisions.

\subsection{Experimental Setup.}

Our implementation is written in the Rust programming language and is available in our public GitHub repository\footnote{\url{https://github.com/Felerius/findminhs}} along with all datasets, logs, run results, and evaluation scripts.
%It uses both the Rust standard library as well as the rand\_pcg library~\cite{randpcg}, an implementation of the PCG family of non-cryptographic random number generators~\cite{o2014pcg}.
%Several other libraries are used for auxiliary purposes like parameter parsing and writing the results.
All auxiliary packages, including the versions used, are listed in the repository as part of the Cargo project format.
For the evaluation, we used version 1.53.0 of the Rust compiler with link-time optimization enabled.
%and using the x86\_64-unknown-linux-gnu target.
The experiments were run on a Gigabyte R282-Z93 (rev.~100) server at 2.6GHz base speed with 1024GB DDR4 (3200MHz) memory.
Runs had a timeout of 24h.

%\paragraph*{Datasets}

We use instances from four sources.

\begin{description}
  \item[\texttt{UCC~\cite{Birnick2020}}]
    contains 134 instances, two for each of 67 databases.  In the first type of instance, the hitting sets correspond to the unique column combinations of the database.  The second type of instances are the transversal hypergraphs of the first type.
  \item[\texttt{CVD~\cite{Bevern2020}}]
    contains cluster vertex deletion instances, derived from weighted graphs of protein similarities~\cite{Rahmann2007,Boecker2007}.
    In the reduction step from weighted graphs to unweighted graphs, we use all edges with non-negative weights.
    This is consistent with the code linked in the paper of Bevern and Smirnov~\cite{Bevern2020}, but differs from the statement in the paper itself, which only uses edges with positive weights.
    Like the authors, we restict us to hypergraphs with at most $10^6$ edges, resulting in a total of 3952 instances.
  \item[\texttt{EN1~\cite{Murakami2014}}]
    contains 159 instances that were previously used to evaluate algorithms for enumerating minimal hitting sets.
    These contain several classes of instances, including real-world and generated instances.
    The original data set contains 172 instances of which we omitted 13 whose size $||\hypergraph||$ exceeds $3 \cdot 10^7$, as the RAM required to run experiments on them proved to be prohibitive.
  \item[\texttt{EN2~\cite{Gainer2017}}]
    contains eleven additional instances that have been used to evaluate enumeration algorithms.  Five of them are derived from metabolic reaction networks and six from interventions in cell signaling networks.
\end{description}

We distinguish between the randomly generated instances (\texttt{rnd}) from the \texttt{EN1} dataset and application specific instances (\texttt{appl}) due to their different structure.
The instances displayed in the various plots are filtered depending on the context.
Figure~\ref{fig:vs-gurobi} includes all $4256$ instances. 
Subsequent plots are restricted to the 136 instances (58 \texttt{rnd}, 78 \texttt{appl}) that finish in 24 hours (excludes 6) and are non-trivial (excludes 4114), that is, instances where our solver runs at least one second in its default configuration.
For experiments that compare different configurations, only the instances finishing in \emph{all} configurations are used.
Effected by this is Figure~\ref{fig:search-space-by-bound} where three instances were dropped due to timeout in some configuration.
Additionally, ten instances where dropped in Figure~\ref{fig:forced-vertices} because they had no forced vertices and two instances where dropped in Figure~\ref{fig:bound-comparison} because they never had a branch pruned due to bounds.

% We report on the dropped instances appropriately.
% 6 did not finish with default
% search-space-by bound: dropped 3 (no finish) + 12 (no branching by default) 
% costly discard find k: nothing dropped
% compare different greedy: nothing dropped
% forced vertices: dopped 10 because no forced verts at al
% which break by bound: dropped 2 because no break by bound in default config

\subsection{Runtime.}
\label{sec:runtime}

Figure~\ref{fig:vs-gurobi} shows the run time of our solver in comparison with Gurobi.
Gurobi is at version 9.1.2, restricted to a single thread, and without a memory restriction.
We note that there are instances where Gurobi uses almost 50GB of memory.
Following Caprara~\cite{Caprara2000}, instances were reduced with the domination rules before running Gurobi on them.
The reduction process is included in the reported run times.
Preliminary experiments showed this to be faster than running the ILP alone.

Our solver is significantly faster than Gurobi on non-random instances; on three quarter of non-random instances at least one order of magnitude.
Contrasting this, there are only three instances (random and non-random) where Gurobi is faster by more than a factor of 4.
Run times for random instances are competitive. 
Gurobi is approximately 1.5 times faster on smaller instances while we are consistently faster on instances that take more than 30 minutes to solve. 
In total, there are 8 instances that finish in the timeout only for our solver. 
There are 2 instances that finish just for Gurobi.

Figure~\ref{fig:operations} shows the fraction of run time that is spent in each step of the algorithm.
As expected from the asymptotic considerations in Section~\ref{sec:algo}, the domination rules dominate the run time although they are executed last and thus avoided when possible.
Random instances spent most time in edge domination since they have many edges and few vertices. 
Non-random instances spend most time in vertex domination.
Still, when taking both classes of instances together, the total time spent is spread over different subroutines and, in the median, no individual task takes more than 20\% of the total time.
Greedy, packing lower bound, and the costly discard repack reduction rule show comparable times.
Although, the repack reduction essentially computes three packings, it is processed later in the loop than the packing lower bound, which explains why the rule does not take three times as much time as the packing.
Finally note that the column for \emph{other} is vanishingly small.
It includes, e.g.,~the instance manipulation and rollbacks as well as logging, timing, and branching.

\subsection{Solver Details.}
\label{sec:solver-details}

In this section we discuss the effectiveness of lower bounds, upper bounds, and reduction rules in detail. 
Moreover, we provide experimental grounds to argue in favor of our choices regarding the solver configuration. 
Specifically, we evaluate the set and order of used lower bounds, the number of checked vertices in the costly discard repack rule, the frequency of greedy invocations, and the order in which reductions are applied.

\paragraph{Lower Bound Effectiveness.}

\begin{figure}
  \includegraphics[width=\linewidth]{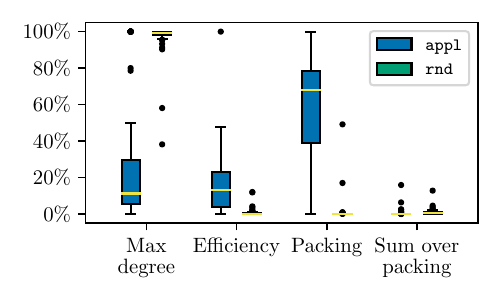}
  \caption{For each instance, the number breaks from the reduction loop split by the responsible bound.
  Values are relative to the total number of breaks for the instance.}
  \label{fig:bound-comparison}
\end{figure}
\begin{figure*}
  \includegraphics[width=\textwidth]{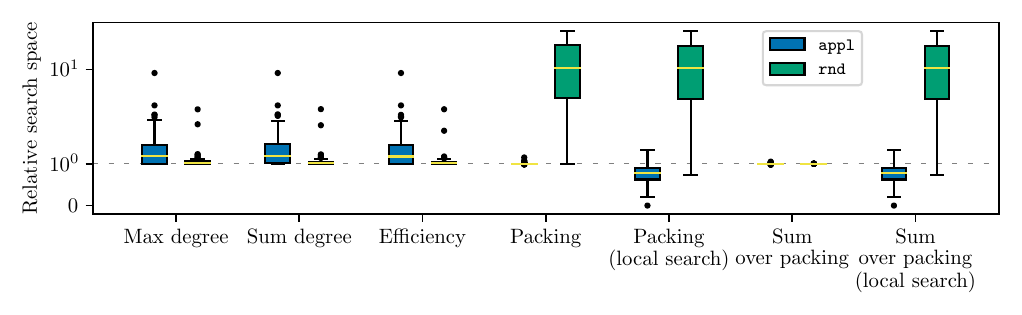}
  \caption{Search space when only using a certain bound, relative to default settings.
  Note the logarithmic y-axis with special handling for zero.}
  \label{fig:search-space-by-bound}
\end{figure*}

Figure~\ref{fig:bound-comparison} counts how often each bound was responsible for pruning a branch in the search tree. 
For random instances, the max-degree bound is sufficient and is responsible for almost all prunes.
Application instances make use of several bounds.
Max-degree still helps but, excluding a few instances, accounts for less than 30\%.
The efficiency and the packing bounds then catch branches that are missed by the previous bounds with the packing bound being significantly more successful than the efficiency bound despite it being run after it.

For the efficiency bound, the outlier at 100\% is due to an instance with a trivial search space of one node that is pruned by the efficiency bound.
Still, there are a few instances where the max-degree bound fails often while the efficiency bound is responsible for 40\% to 60\% of all prunes.

There are even more instances which almost exclusively depend on the packing bound.
This can be explained due to the packing bound representing a different approach to obtain the lower bound than the max-degree or efficiency bound and thus performs well on a different kind of instances.  The max-degree bound performs well if the high degree vertices do not share many edges, i.e., if the instance can be covered by selecting few high-degree vertices.  On the other hand, many edges containing multiple high-degree vertices makes it possible to have many edges containing only vertices of lower degree, which facilitates large packings.

Regarding the last bound, note that the sum-over-packing bound rarely applies. 
However, Figure~\ref{fig:operations} shows that its run time is negligible since the previously computed packing is reused. 
For some instances the bound actually prunes a significant number of branches.

With the question answered to what extent our chosen lower bounds contribute to pruning, it remains to show that it is their combination and not one bound alone that is responsible for the overall performance.
Figure~\ref{fig:search-space-by-bound} shows the relative search space when using only one bound compared to using our default configuration for the solver.
Max-degree, sum-degree, and efficiency all behave similar, that is worse than with all bounds.
On the other hand, for random instances only using the packing bound leads to significantly higher search space.
These instances, however, are solved easily when using the sum-over-packing instead of the packing alone.
In fact, using only sum-over-packing is almost always as good as using the combination of bounds.  Nonetheless, it still makes sense to use the other bounds before: The packing bound is free as we have to compute a packing anyway to apply sum-over-packing.  Moreover, the simpler bounds can be computed more quickly than a packing (recall Figure~\ref{fig:operations}) but are often sufficient as can be seen in Figure~\ref{fig:bound-comparison}.

As mentioned in Section~\ref{sec:packing-computation}, we implemented versions of the packing with local search and included them in Figure~\ref{fig:search-space-by-bound}.
In the following we consider the influence of adding local search to packing or sum-over-packing.
For application instances, adding the local search to the packer or sum-over-packing bound results in a slight improvement over the combination of all other bounds.
However, preliminary testing showed local search to be too computationally expensive to justify the reduction in search space it yields.
On random instances, adding local search to packing does not change the search space, while adding it to sum-over-packing surprisingly increases the search space.
Further investigation revealed that a packing that was augmented with local search is enlarged to the point that taking the highest degree vertex of each edge in the packing constitutes for enough total degree to cover the whole instance.  In this case, the sum-over-packing bound is equal to the packing bound.

\paragraph{Upper Bound Effectiveness.}
The greedy upper bound is used to initialize and, during a run, improve the best solution found so far.
Still, the solver could be implemented without greedy at all.
The upper bound would then be initialized to contain all vertices in $V$ and updated when the branching reaches a hitting set.
Running the greedy subroutine has the benefit that it helps to find solutions before reaching the associated leaf in the search tree and thus facilitates earlier pruning.
In the following, we compare four different frequencies of running greedy to recompute the upper bound.
The alternatives are to not use greedy at all, run greedy once before the reduction loop (which is what we do in the final configuration of our algorithm), run greedy every iteration of the loop (i.e., as item zero in Section~\ref{sec:algo-order}), or to run it every loop just before the expensive reductions (i.e., between items 7 and 8 in Section~\ref{sec:algo-order}).

\begin{figure}
  \includegraphics[width=\linewidth]{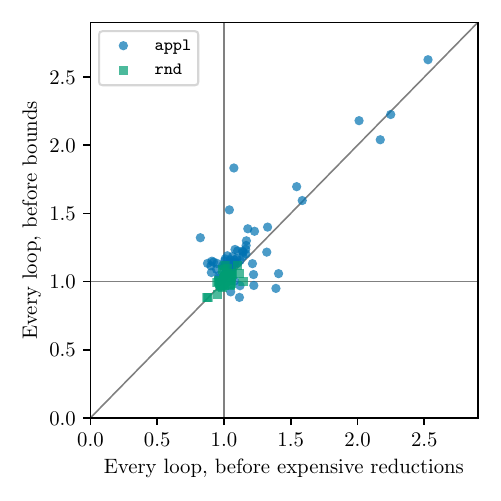}
  \caption{Run times of two non-default greedy modes relative to the run time when using greedy once per node in the search tree. Each instance is a data point.}
  \label{fig:where-greedy}
\end{figure}

Figure~\ref{fig:where-greedy} compares the latter three alternatives.
The baseline in the plot (and default configuration for the solver) is to run it once before the loop. 
The axes show the relative run time compared to this baseline.
A point in the lower left quadrant means that the baseline is the worst out of the three;
a point in the upper right quadrant that it is best, while the rest ranks it in the middle. 
Samples are concentrated in the center but slightly tilted to the upper right.
Additionally, there are no outliers that favor running greedy every loop while there are outliers that heavily slow down when deviating from our baseline.

\begin{figure}
  \includegraphics[width=\linewidth]{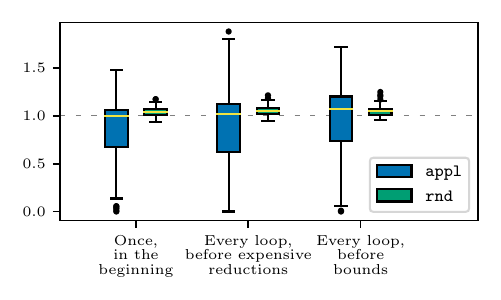}
  \caption{Relative run time for all three greedy modes compared to not using greedy upper bounds.}
  \label{fig:greedy-vs-off}
\end{figure}

Figure~\ref{fig:greedy-vs-off} shows run times of all alternatives that use greedy compared to not using greedy at all, that is, initializing the upper bound to $V$ and find better solutions only in leaf nodes while branching.
Surprisingly, the median favors not using greedy at all but the outliers are heavily in favor of using greedy.
Running greedy once before the reduction loop is never slower than 1.5 times the run time of no greedy, 1.02 times slower in the median, and up to 600 times faster at best, which makes it a good choice for the final configuration of our algorithm.
Note that the benefits of greedy are restricted to application instances.
On random instances there are no heavy outliers in favor of any configuration.

\paragraph{Reduction Effectiveness.}

\begin{figure}
  \includegraphics[width=\linewidth]{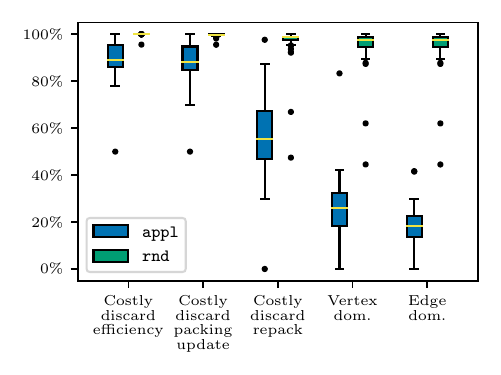}
  \caption{
    For each instance, the number of reduction loop runs that reach a given reduction.
    Values are relative to all runs that do not break due to bounds.
    The forced vertex rule is excluded as it is the first and thus always reached.
  }
  \label{fig:reduction-runs}
\end{figure}

Before we establish how effective each specific reduction is, we first investigate how often each reduction is actually reached in the reduction loop as shown in Figure~\ref{fig:reduction-runs}.
Differences between two adjacent reductions express the success rate of the left one.
Random instances, again, behave completely different than application specific instances.
They almost always execute all reductions, because all reductions before the last one are unsuccessful.
For application instances, all rules contribute somewhat.
Although the domination rules have the highest run time share (recall Figure~\ref{fig:operations}) they are only executed in 20--30\% of loop iterations for most instances.
The most frequent end to an iteration are a successful costly discard rule with packing updates or repacking. 
These rules are reached in more than 80\% of iterations for most instances and the next rule (vertex domination) is checked 30\% of the time in the median.
Note that an iteration only ends in the repack step if the costly discard rule was unsuccessfully checked with a packing update before succeeding through a repack,  giving evidence to the usefulness of repacking.

\begin{figure}
  \includegraphics[width=\linewidth]{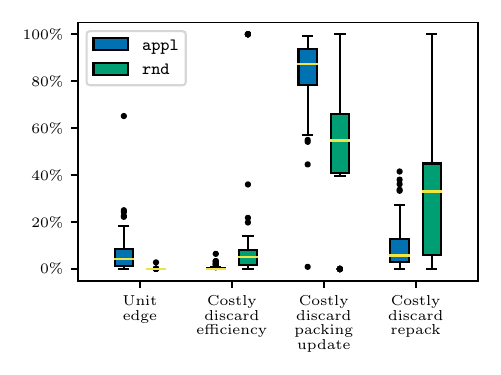}
  \caption{For each instance, the number of found forced vertices by reduction rule. 
  Values are relative to the total number of forced vertices for the instance.}
  \label{fig:forced-vertices}
\end{figure}

\begin{figure*}
  \includegraphics[width=\textwidth]{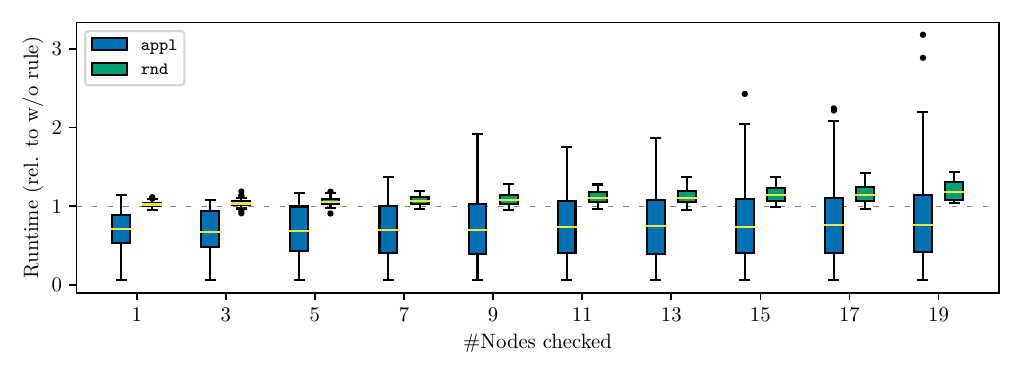}
  \caption{Run time when using different settings for the costly discard packing from scratch rule. All values are relative to the run time when that rule is disabled.}
  \label{fig:from-scratch}
\end{figure*}

Figure~\ref{fig:forced-vertices} compares the effectiveness of the reductions that force a vertex to be included in the solution.
The effectiveness is measured by the number of forced vertices.
The unit edge rule and the costly discard rule with the efficiency bound catch almost no vertices compared to the two packing rules.
Since they are very easy to compute and find a good amount of application on some instances they are nonetheless worth trying.
Above 80\% of forced vertices in application instances are found with the costly discard packing update rule.
This constitutes no contradiction to Figure~\ref{fig:reduction-runs} because, although the repack rule is applicable as often, the packing update rule can find multiple forced vertices at once.
The fact that the repack rule still finds a large amount of all forced vertices while being applied only after packing update failed to find anything, again, emphasizes that the updated packings are not as good as the packings that are constructed from scratch.
The results for random instances are less expressive because only a small number of reductions are applicable, as established in the previous paragraph.
Nevertheless, they follow the same trend as the other instances but rate packing update lower and packing repack higher.

One degree of freedom for the repacking is the number of vertices for which we apply it.  Recall from Section~\ref{sec:algo-order} that we repack for the $c = 3$ vertices of maximum degree in our final configuration.  
Figure~\ref{fig:from-scratch} shows the run time for different values of $c$ relative to the run time when not repacking.
One can make two main observations.  
First, for random instances, the cost of repacking outweighs the gain leading to slightly increasing median run times for increasing $c$.
Second, repacking helps significantly on non-random instances but there is no additional gain in repacking more than three times, which has the lowest median.  
Thus, by using $c = 3$, we obtain a good balance between increasing run time for random instances only slightly while obtaining big speedups for some application instances.

%%% Local Variables:
%%% mode: latex
%%% TeX-master: "../paper"
%%% End:

\section{Conclusion}

We provide a fast branch-and-bound solver that beats a modern ILP solver, which is the state-of-the-art for solving the minimum hitting set problem.
Our implementation provides a baseline for future work in this direction.
We explain the basic building blocks of our algorithm --- which are lower bounds, upper bounds, and reduction rules --- 
and experimentally evaluate their run time and efficiency to find a good configuration of used rules and bounds.
We confirm the effectiveness of Weihes reduction rules noted in previous works.
Another crucial part of the algorithm turns out to be the quality of lower bounds.
The parameter-dependent \emph{Costly Discard Rule} builds upon lower and upper bounds and contributes significantly to the performance of our algorithm.
Lastly, we find that the algorithm behaves differently on random inputs.
In the future, it would be interesting to determine why non-random instances are easier for our solver than random instances and if their structure can be exploited to design even faster algorithms for practical instances.
 
\bibliography{ms}

\clearpage

\appendix

\section{Additional Evaluation Results}

This section contains additional evaluation results that have been cut from the main paper due to space constraints.

\subsection{Search Space and Bounds.}
\label{sec:search-space-bounds}

\begin{figure*}
  \includegraphics[width=\textwidth]{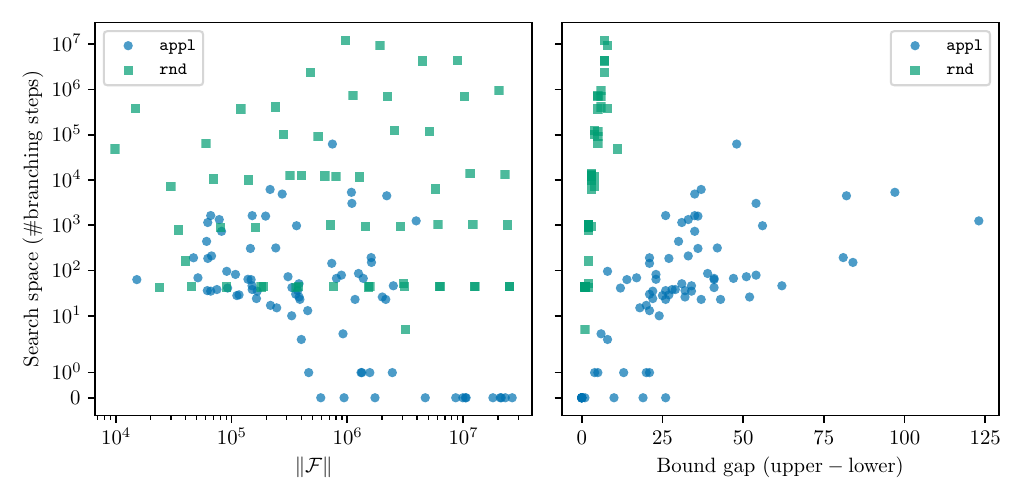}
\caption{Search space compared to instance size (left) and difference between upper and lower bound (right). Each instance represents one sample.
The y-axis is logarithmic with special handling for zero.}
\label{fig:search-space}
\end{figure*}

Figure~\ref{fig:search-space} shows the size of the search space depending on the instance size and the difference between the initial upper and lower bound.
The lack of samples with low search space and instance size is an artifact due to the exclusion of instances that are solved in less than a second.

There appears to be no clear correlation to indicating that instance size reflect difficulty.
On the other hand, the bound gap is usually a good estimate for the difficulty of an instance to a given branch-and-bound solver, because the solver lowers this gap during execution and is finished if the difference reaches zero. 
The random instances exhibit a distinct exponential growth in search space with growing gap.

\begin{figure}
  \includegraphics[width=\linewidth]{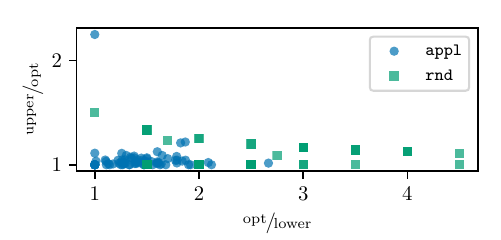}
  \caption{The plot shows the initial lower bound relative to opt on the x-axis and the upper bound relative to opt on the y-axis. Each instance represents one sample.}
  \label{fig:bound-gaps}
\end{figure}

Figure~\ref{fig:bound-gaps} further explores the difference between lower and upper bounds.
The initial upper bounds are already close to the optimum with all but three being less than 1.3 times the optimum.
Lower bounds are spread out more.
Some instances have lower bounds that are not even a third of the optimum.
Interestingly, the instance with with worst upper bound has a perfect lower bound.

\begin{figure}
  \includegraphics[width=\linewidth]{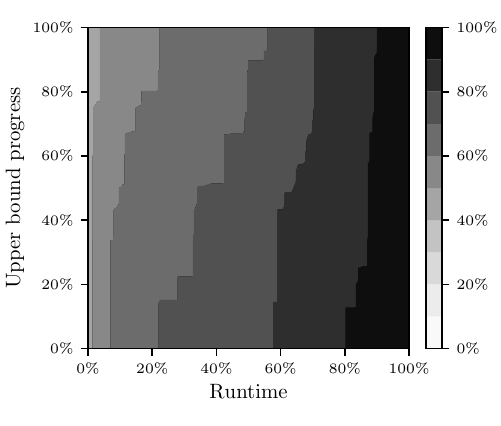}
  \caption{For a given run time and a given amount of progress from the initial upper bound towards opt, how many instances have reached that progress at that time.}
  \label{fig:bound-updates}
\end{figure}

Figure~\ref{fig:bound-updates} shows the progression and gradual lowering of the upper bound during execution.
Each individual instance starts in the lower left corner at (0,0) and progresses to the upper right corner at (100,100).
Under the assumption that this progression is linear we would get white above and black below the main diagonal.
However, the plot can be better described as growing darker from left to right.
This verticality means that during a run, upper bounds progress is often achieved all at once.
Also the upper bound is surprisingly good with more than 40\% of the instances already having a perfect upper bound to start with.
After 25\% of the run time the optimum has been found (100\% progress) for more than 60\% of instances.  The remaining 75\% of the run time are used for proving its optimality.

\subsection{Performance on Geometric Inhomogeneous Random Graphs.}
To obtain realistic instances with controllable structural properties, we use geometric inhomogeneous random graphs (GIRGs)~\cite{Bringmann2019}, which are a generalization of hyperbolic random graphs~\cite{Krioukov2010}.
This lets us isolate the effect of a single structural property on the algorithm performance while keeping most other properties of the instance fixed.
We discuss --- in this order --- the degree distribution, the hyperedge size distribution, the ratio of vertices to hyperedges, the average hyperedge size, the instance size, as well as the amount of clustering.

\paragraph*{Generator Setup.}
We use the efficient generator by Bl\"{a}sius et al.~\cite{blaesius2019efficiently} and modify it to generate bipartite graphs which can be interpreted as hitting set instances.
The modified generator allows to specify positions and weights for the two partitions separately.
Since we cannot use the provided method to estimate the scaling constant which controls the average degree we perform a binary search instead.
Another problem is that the generator yields hyperedges whose size is controlled \emph{in expectation}.
Since hyperedges of size zero make the instance trivially unsolvable, we discard empty hyperedges.
The discard is done after the binary search.
Therefore the degree is sometimes slightly larger than intended.
We monitor the amount of discarded edges to (manually) guarantee that only a very small fraction is discarded.
The modified generator, plotting code, raw results, as well as execution logs are publicly available\footnote{\url{https://github.com/chistopher/sat-girgs}}.

Unless noted otherwise, the hypergraphs have 200 vertices, 200 hyperedges, and an average hyperedge size of 10.
We set the internal parameters of the GIRG model to temperature 0, dimension 2, and a power-law exponent of 2.8.
Each box of the box plots summarizes 50 graphs generated with the same set of parameters but different seeds.
%After investigation the distributions, we repeated the subsequent experiments for all three distribution combinations with at least one heterogeneous distribution.
%To mirror the application instances, we report the results for homogeneous hyperedge sizes, heterogeneous vertex degrees and note if the other combinations behave differently.

\paragraph*{Degree Distributions.}
\begin{figure}
\includegraphics[width=\columnwidth]{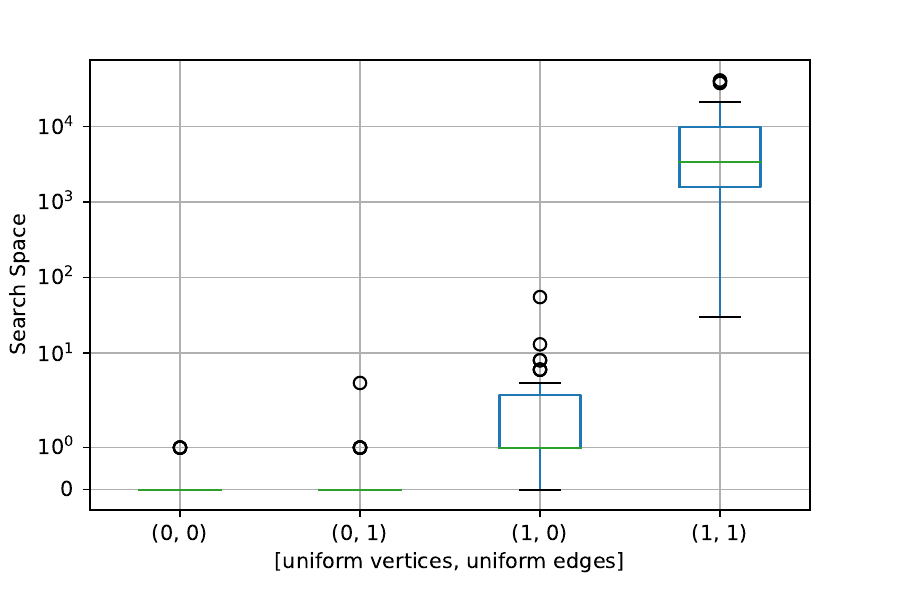}
\caption{Search space of the solver when choosing a homogeneous or heterogeneous distribution for vertex degrees and hyperedge sizes, respectively.}
\label{fig:girgs-dist}
\end{figure}

Vertex degrees and hyperedge sizes are two degrees of freedom when generating instances. 
Viewed as a bipartite graph, these correspond to the degree distributions of the two partitions.
While there are numerous reasonable possibilities for the distributions, we focus on homogeneity vs heterogeneity represented by a Poisson binomial and a power-law distribution, respectively.
Figure~\ref{fig:girgs-dist} shows the search space of the solver for all four combinations.
Heterogeneity drastically reduces the search space.
Moreover, the instances with just a heterogeneous vertex degree are slightly easier than the ones with just heterogeneous hyperedge sizes.
This fits nicely with the data sets from the previous section, where most application instances have homogeneous hyperedge sizes and heterogeneous vertex degrees.

\paragraph*{Vertex to Edge Ratio.}
\begin{figure}
\includegraphics[width=\columnwidth]{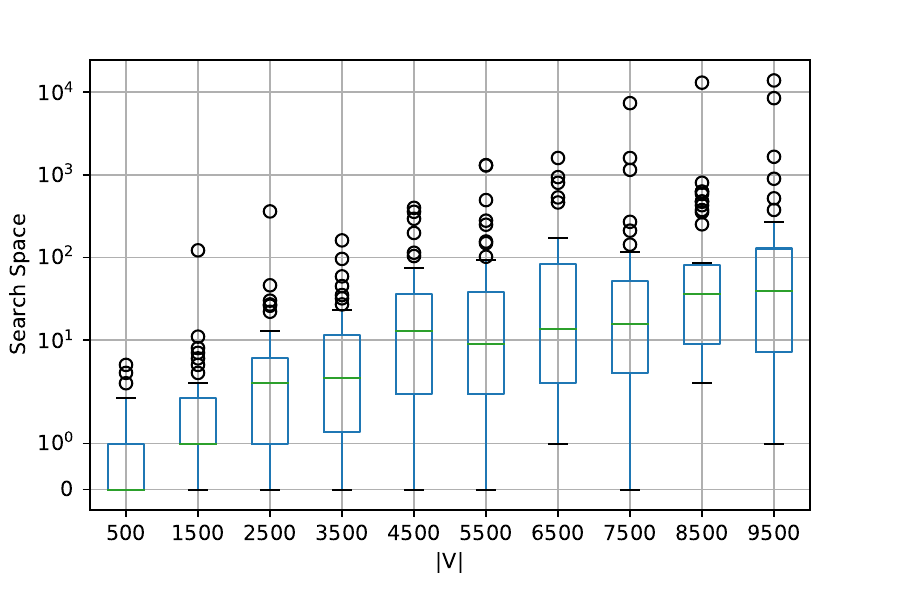}
\caption{Search space of the solver for growing number of vertices and 5000 hyperedges. Vertex degrees are heterogeneous and hyperedge sizes homogeneous.}
\label{fig:girgs-ratio}
\end{figure}
Figure~\ref{fig:girgs-ratio} considers the effect of the ratio between vertices and edges on performance.
We focus on homogeneous hyperedge sizes and heterogeneous vertex degrees to mirror the application instances.
Note that the instance size is significantly larger than in Figure~\ref{fig:girgs-dist}.
With this choice of distribution, most instances would be trivial otherwise.
The figure shows the search space for instances with 5000 hyperedges with average size of ten over a growing number of vertices.
The figure indicates the trend that having a large amount of vertices in comparison to the number of hyperedges makes the instances more difficult.
Aside from the larger input size which should not matter as much, the trend can be explained by the fixed hyperedge size.
Since the average hyperedge size is fixed to ten, more vertices implies smaller vertex degrees which in turn implies larger solution sizes.
Large solutions are generally difficult to find, since they require either more successful reductions or more branching.
Preliminary experiments confirmed a similar trend for other distributions as long as at least one of them is heterogeneous.

\paragraph*{Hyperedge Size.}
\begin{figure*}
\centering
\begin{subfigure}[b]{0.49\textwidth}
\includegraphics[width=\columnwidth]{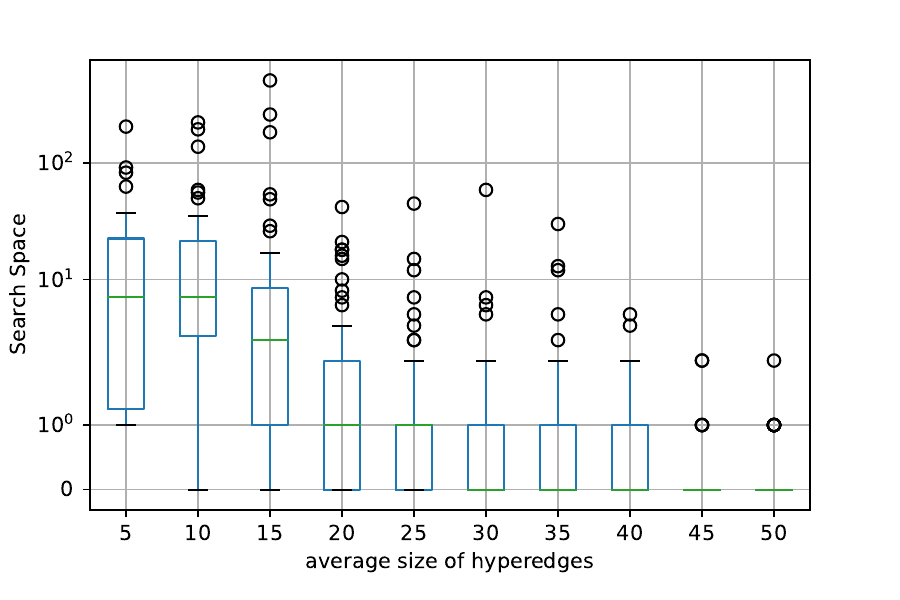}
\caption{Search space for heterogeneous vertex degree.}
\label{fig:girgs-deg-unicls}
\end{subfigure}
\begin{subfigure}[b]{0.49\textwidth}
\includegraphics[width=\columnwidth]{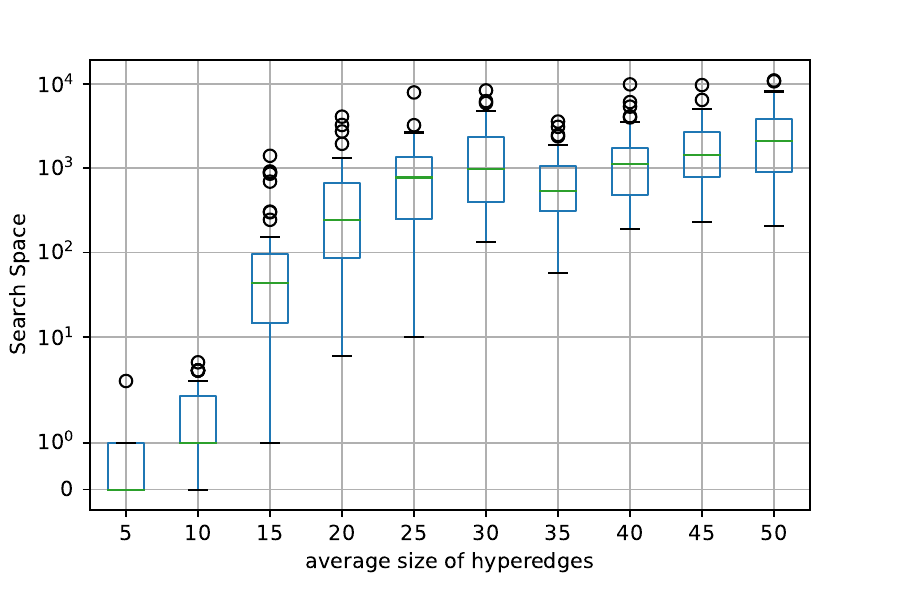}
\caption{Search space for heterogeneous edge size.}
\label{fig:girgs-deg-univar}
\end{subfigure}
\caption{Search space by hyperedge size. In (a) $|V|=5000, |\mathcal{F}|=5000$. In (b) $|V|=200, |\mathcal{F}|=200$.}
\label{fig:girgs-deg}
\end{figure*}

The extrema in hyperedge size are both easy. 
Too small hyperedges of size zero or one lead to trivial instances while too large hyperedges lead to tiny (one or two vertex) solutions.
Figure~\ref{fig:girgs-deg} shows the search space over a growing average size of hyperedges.
For heterogeneous vertex degrees larger hyperedges make the instance easier (see Figure~\ref{fig:girgs-deg-unicls}).
In contrast, larger hyperedges make the instance harder for heterogeneous hyperedge sizes (see Figure~\ref{fig:girgs-deg-univar}).
The former is due to the trivial one or two vertex solutions that are reached far earlier when vertex degrees are heterogeneous.
We explain the latter by the domination rules which are less effective the larger the hyperedges.

\paragraph*{Clustering.}
\begin{figure}
\includegraphics[width=\columnwidth]{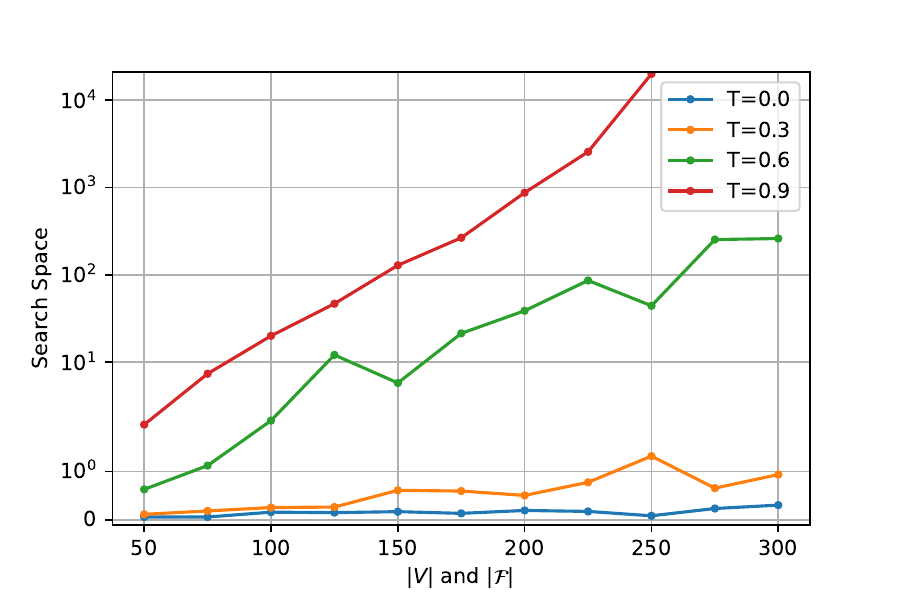}
\caption{Search space by instance size for different amounts of clustering. Vertex degrees are heterogeneous and hyperedge sizes homogeneous.}
\label{fig:girgs-temp}
\end{figure}
The temperature parameter $T$ of the GIRG model controls the clustering, that is, it controls the extent to which the geometry is respected during generation.
Higher values of $T$ means less clustering.
Figure~\ref{fig:girgs-temp} shows the search space over growing instance size for different values of $T$.
Due to high variance this plot is averaged over 200 repetitions per data point instead of 50.
As the algorithm was built for realistic instances, which usually have a high amount of clustering, it is not surprising that the search space grows rapidly when there is less clustering.
A lack of clustering not only makes the instance harder but higher $T$ also makes the solver scale worse with instance size.
We suspect the domination rules to be the reason for this drastic impact on performance because they benefit from high clustering.

\end{document}